\documentclass[modern]{aastex62}
\usepackage{amssymb,epsfig,subfigure,graphicx,makecell,color}
\usepackage[normalem]{ulem}

\newcommand{\msun}{$\rm M_\odot$}

\newcommand{\lsun}{$\rm L_\odot$}

\newcommand{\kms}{$\rm km\,s^{-1}$}
\newcommand{\hii}{H{\sc ii}}

\newcommand{\nirspec}{{\small NIRSPEC}}
\newcommand{\nirspao}{{\small NIRSPAO}}
\newcommand{\scam}{{\small SCAM}}

\newcommand{\alma}{{\small ALMA}}

\newcommand{\co}{{\small CO}}
\newcommand{\fwhm}{{\small FWHM}}

\newcommand{\ngc}{{\small NGC}$\,$5253}

\newcommand{\bra}{Br$\,\alpha$} 
\newcommand{\ha}{H$\,\alpha$}
\newcommand{\hei}{He~{\sc i}}
\newcommand{\degree}{$^{\circ}$}
\newcommand{\uchii}{UCH{\sc ii}s}

\accepted{April 26, 2018}
\submitjournal{ApJ}

\shorttitle{AO Observations of the Supernebula in NGC~5253}
\shortauthors{Cohen et al.}

\begin{document}

\title{%
Ionized Gas Motions and the Structure of Feedback Near a Forming Globular Cluster in NGC~5253 
}

\correspondingauthor{Daniel Cohen}
\email{dcohen@astro.ucla.edu}
 
\author{Daniel P. Cohen}
\affiliation{University of California, Los Angeles, CA 90095-1547, USA}

\author{Jean  L. Turner}
\affiliation{University of California, Los Angeles, CA 90095-1547, USA}

\author{S. Michelle Consiglio}
\affiliation{University of California, Los Angeles, CA 90095-1547, USA}

\author{Emily C. Martin}
\affiliation{University of California, Los Angeles, CA 90095-1547, USA}

\author{Sara C. Beck}
\affiliation{School of Physics and Astronomy, Tel Aviv University, Ramat Aviv, Israel}


\begin{abstract}
We observed Brackett $\alpha$ 4.05$\mu$m emission 
towards the supernebula in {\ngc} with 
 \nirspec\ on Keck II
in adaptive optics mode, \nirspao, to probe feedback from its exciting embedded super star cluster (SSC).
\nirspec 's Slit-Viewing Camera was simultaneously 
used to image the 
K-band continuum at $\sim$0\farcs1 resolution. 
 We register the IR continuum with HST imaging, 
 and find that the 
 visible clusters  
 are offset from the K-band 
peak, which coincides with the \bra\ peak 
of the supernebula and its associated molecular cloud.
The spectra of the supernebula exhibit \bra\ emission with a strong, narrow core. 
The linewidths are 65--76~\kms, \fwhm, 
 comparable to those around individual ultra-compact \hii\ regions within our Galaxy.  
A weak, broad (\fwhm$\simeq$150-175 {\kms}) component is detected on
 the base of the line, which could trace a population of sources with high-velocity winds.
The core velocity of \bra\ emission 
shifts by +13 {\kms} from NE to SW across the supernebula,
 possibly indicating a bipolar outflow from an embedded
object, or linked to a foreground redshifted gas filament. 
The results can be explained if the supernebula comprises 
thousands of ionized wind regions around individual massive stars, stalled in their expansion due to critical radiative cooling and unable to merge to drive
a coherent cluster wind.
 Based on the absence of an outflow with large mass loss,
 we conclude that feedback is currently ineffective at 
dispersing gas, and the SSC retains enriched material out of which 
it may continue to form stars.

\end{abstract}

\keywords{%
galaxies: dwarf ---
galaxies: individual (NGC~5253) --- 
galaxies: starburst --- 
galaxies: star clusters: general --- 
galaxies: star formation --- 
ISM: HII regions  
}

\section{INTRODUCTION} \label{sec:intro}

Massive star clusters are fundamental components of galactic evolution.
Details of gas retention or expulsion 
during cluster formation must be reconciled with the existence of long-lived massive clusters, globular clusters (GCs), 
and with multiple stellar populations observed in the GCs 
 \citep{piotto2015,silich2017}. Massive clusters  
 contain concentrations of the most massive stars \citep{kroupa2002,
kroupa2013}, whose evolution can potentially induce 
great radiative, dynamical, and chemical changes in their 
host galaxies.
To probe the formation 
of GCs, we can utilize  
observations of local protoglobular cluster candidates, super star clusters (SSCs). 
They can be as young as $\sim$1-10 Myr \citep[e.g.,][]{turner2009}.

One of the most thoroughly studied forming GC is the compact radio/IR nebula, 
the ``supernebula", in the 
dwarf starburst galaxy {\ngc}. 
First discovered in the radio from its compact, thermal free-free emission \citep{beck1996}, this giant \hii\ region 
\citep{beck1996, calzetti1997, turner1998}
 is powered by a massive SSC ($M_{\rm vir}=2.5\times10^5$ {\msun}) with $L\sim 5\times 10^8$ {\lsun} and 
$\sim$2000 O stars 
\citep{turner2000,meier2002,turner2003,vanzi2004,hunt2005,calzetti2015,smith2016,turner2017}. The supernebula may be part of the core of a larger star-forming region containing 4000-7000 O stars \citep{meier2002}. 
What we refer to as the supernebula is its $\sim$3 pc  core \citep{turner2004,bendo2017,consiglio2017}.

The supernebula is still deeply embedded within its natal cloud. Extinction from dust is so high that it 
is invisible in wavelengths shorter than the NIR,
and is the brightest source in the galaxy at wavelengths longer than the $H$-band \citep{turner2003,herrero2004,martin2005}. Its dustiness ($A_V\gtrsim 16-18$), together with detection of Wolf-Rayet signatures \citep{conti1991,schaerer1997,sanchez2007,ibero2010,westmoquette2013}, suggest the presence of stars $\sim$1-3 Myr in age \citep[e.g.,][]{herrero2004,calzetti2015,
smith2016}. 
Submillimeter observations 
found a warm \co($J$=3--2) cloud, 
Cloud ``D1", that is coincident to within 0.6 pc of the \hii\ region 
and within 2 \kms\ of IR and radio recombination line velocities \citep{turner2015,turner2017}. 
This gives the first indication that there are potential star-forming molecular clouds within
the cluster itself.
 The \co\ gas is remarkably quiescent ($\sim$22 {\kms} {\fwhm}) considering its  
extreme environment, within a 3 pc region containing thousands of O stars. The 
linewidth of Cloud D1
reflects only the gravitational motions of the cluster, with no evidence for molecular outflow. 
The supernebula thus does not appear to be launching a galactic wind of
the kind seen, for example, in NGC~253 \citep{sakamoto2006,bolatto2013}. 

Observations of the supernebula suggest feedback is similarly ineffective in driving outflows of ionized gas from the exciting SSC. 
\citet{turner2003} observed the Brackett $\alpha$ and $\gamma$ recombination lines in the supernebula
with a 0\farcs5 slit using \nirspec\ and found, in addition to high extinction,
 a relatively small linewidth (\fwhm\ of 76 {\kms}), similar to widths seen in Galactic \hii\ regions and comparable to the escape velocity of the cluster. 
Measurements of mid-IR emission lines \citep{beck2012} and radio recombination lines \citep[RRLs;][]{rico2007,bendo2017} confirm 
this measurement. 
The gravity of the massive cluster clearly has a large role in shaping the gas motions and could potentially launch a cluster wind.
However, numerical 
wind models suggest that if the {\ngc} supernebula is sufficiently
chemically enriched, strong radiative cooling can 
suppress a starburst wind \citep{silich2003,silich2004,silich2017}.
Localized chemical enrichment has in fact been indicated in {\ngc} by 
nuclear abundance studies \citep{walsh1989,kobulnicky1997,sanchez2007,ibero2010,westmoquette2013} and by 
submillimeter continuum emission from dust \citep{turner2015}, and in other SSCs \citep{consiglio2016}.

In this study we present high resolution (0.1{\arcsec}) observations of Br$\,\alpha$ emission across the {\ngc} supernebula, made with the \nirspec\ spectrograph on Keck II using laser-guided adaptive optics 
({\nirspao}).  
At 4 $\mu$m, the \bra\ line is less  
effected 
by extinction than \ha\ or Br$\,\gamma$, allowing us to probe gas embedded within the SSC. 
What is happening to the gas in this potential proto-GC, which appears to be in its infant phase? The goal of this investigation is to use our \bra\ observations to infer the kinematics of the supernebula 
and determine the influence and fine structure of feedback from the cluster. 
At the 3.8 Mpc distance of \ngc\ \citep{sakai2004},  $0.1\arcsec = 1.8$ pc.


\section{OBSERVATIONS AND DATA} \label{sec:data}

The supernebula in \ngc\ was observed with \nirspao\ 
\citep{mclean1998} in the second half-night on May 1, 2015.  
Spectra were taken 
in high-resolution mode (R$\sim$25,000), in the KL band, using the 2.26\arcsec$\times$ 0.068\arcsec\ slit. The echelle and cross-disperser angles were set to 64.5\degree\ and 34.12\degree, respectively, yielding a wavelength coverage of roughly 4.03-4.08 $\mu$m in the 19th echelle order.  The spatial resolution of the observations is $\simeq$0\farcs12, determined from the continuum spectra of several calibration stars.

The  
slit was oriented 
at a position angle of 113.5\degree\ (so that the slit-parallel direction is close to the E-W axis), with each position separated by the slit width ($\simeq$0\farcs07). 
The slit positions are hereafter labelled: N2, N1, S1, S2 from north to south.  
During spectral exposures, the the Slit-Viewing Camera ({\scam}) was used to
simultaneously and continuously image the slit on the sky in 
the K band, 
giving a record of the slit position 
relative to the NIR continuum.

A stacked, reduced \scam\ image was 
constructed to study the spatial distribution of gas and determine registration of slit positions with respect to the radio supernebula. To form this image, raw \scam\ images with the supernebula on-slit were subtracted by images taken with the supernebula off-slit (at the end of the night during sky exposures). Then, for each separate slit position,  
clusters visible in the field-of-view were used to align all sky-subtracted \scam\ images.
The slits were masked out in each of the four images, 
which were then aligned, and the average was computed 
over each pixel to form a combined, slit-free \scam\ image.

  To reduce the \nirspao\ spectra, we 
  applied spatial and spectral rectification to 
  all raw \nirspec\ images using 
  the {\small IDL}-based {\small REDSPEC} pipeline\footnote{\url{https://www2.keck.hawaii.edu/inst/nirspec/redspec.html}}. 
 Arc lamps used for calibration were not functioning properly for the observing run, yielding no
 emission lines in the \bra\ order. New arc lamp spectra obtained on a later date (with the same configuration) detect emission lines in this order, but the resulting wavelength calibration is inaccurate in its wavelength zero-point. The issue can be partially corrected by determining the offset between the rest wavelength of \bra\, and the \bra\ absorption line detected in a calibrator star spectrum. However, there is no measurement of radial velocity for the calibrator. Centroid velocities inferred from the \bra\ emission of the supernebula are $\sim$30-40 {\kms} higher than the previously-determined velocity of $391 \pm 2$ {\kms} (barycentric), measured for the H30$\alpha$ line \citep{bendo2017} and for the narrow component on the [S{\sc iv}] 10.5 $\mu$m line \citep{beck2012}. Fortunately, we have \hei\ and \bra\ lines to determine the wavelength scale, and the measured and expected separation between these lines are consistent to within 2-3\%, or $\lesssim$6 {\kms}.
Because we are only interested in the velocity linewidths and the {\it relative} variation of centroid velocity across the supernebula region, the velocity zero-point does not change any of our results or conclusions. To facilitate comparison with other studies, we shift the spectra so that the \bra\ line in the average 
spectrum across the four slit positions is at a centroid velocity of 391 {\kms}.

Following rectification, individual spectra were 
sky-subtracted and divided by a median-normalized flat-field image, and hot/cold pixels were removed. 
Reduced exposures were median-combined 
to form the the 2D spectrum, or echellogram, for each of the four slit positions. 
Fringe patterns due to interference within the detector are apparent in the echellograms.
We tested 
a correction for fringing but find that it does not significantly change any results or derived quantities. 
Since these are slit spectra made through AO observations
 it is difficult to estimate the absolute calibration of line strength, which has been
done elsewhere \citep{turner2003}.  
The resulting four echellograms 
have spectral resolution of $\sim$12 {\kms} and spatial resolution of $\simeq$0\farcs1 (with pixel size 0\farcs018 pix$^{-1}$ in the spatial direction).

For analysis of kinematics (\S~\ref{sec:spectroscopy}), 1D spectra are extracted 
by summing or averaging rows in the echellograms. Gaussian models are fit
to the emission in the extracted spectra to infer line properties.


\section{Nature of the NIR Continuum and Relation to Visible Clusters}  \label{sec:continuum}

\begin{figure}[!ht]
\centering
\subfigure{\includegraphics[width=0.52\textwidth]{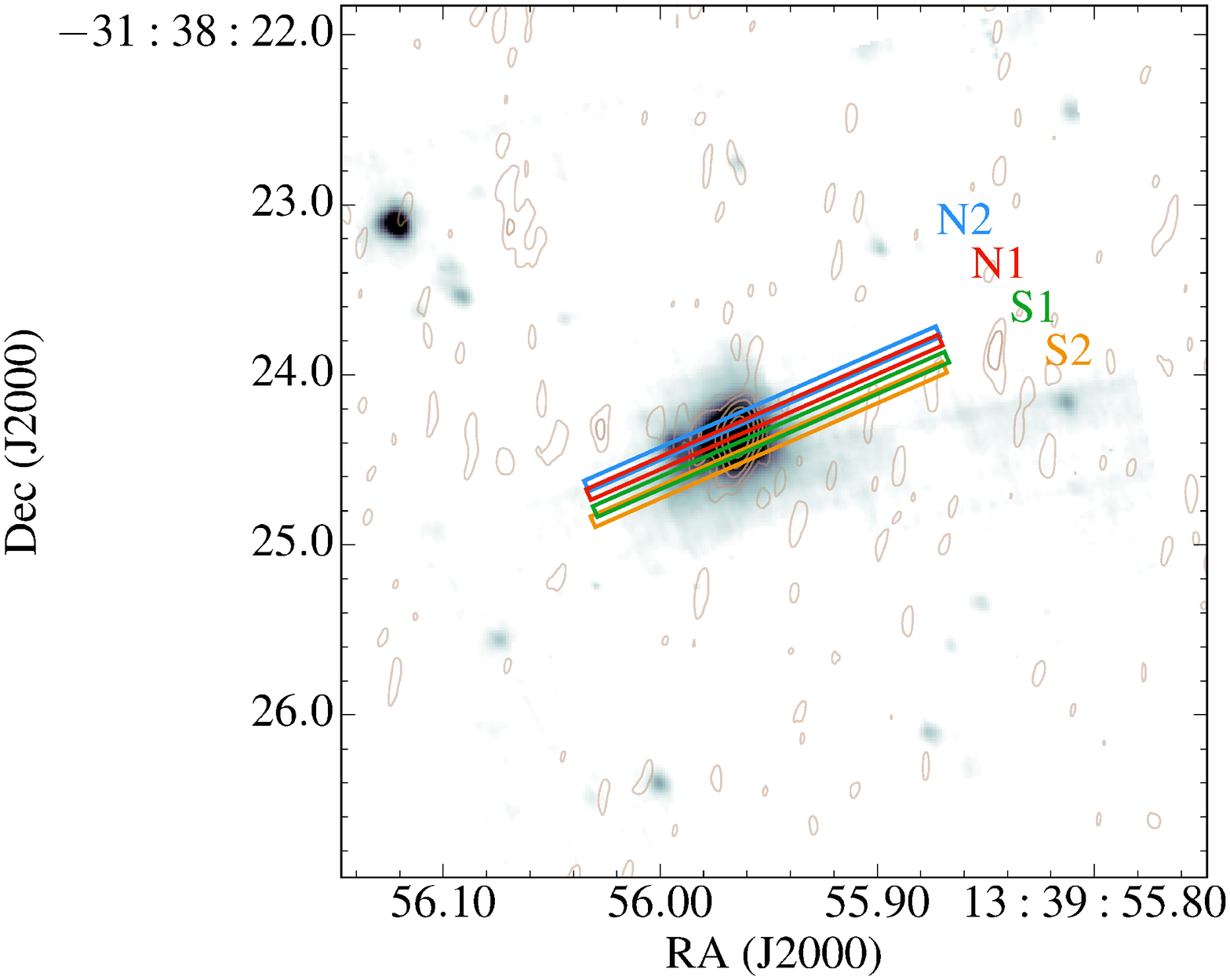}}\hspace{3mm}
\centering
\subfigure{\includegraphics[width=0.35\textwidth]{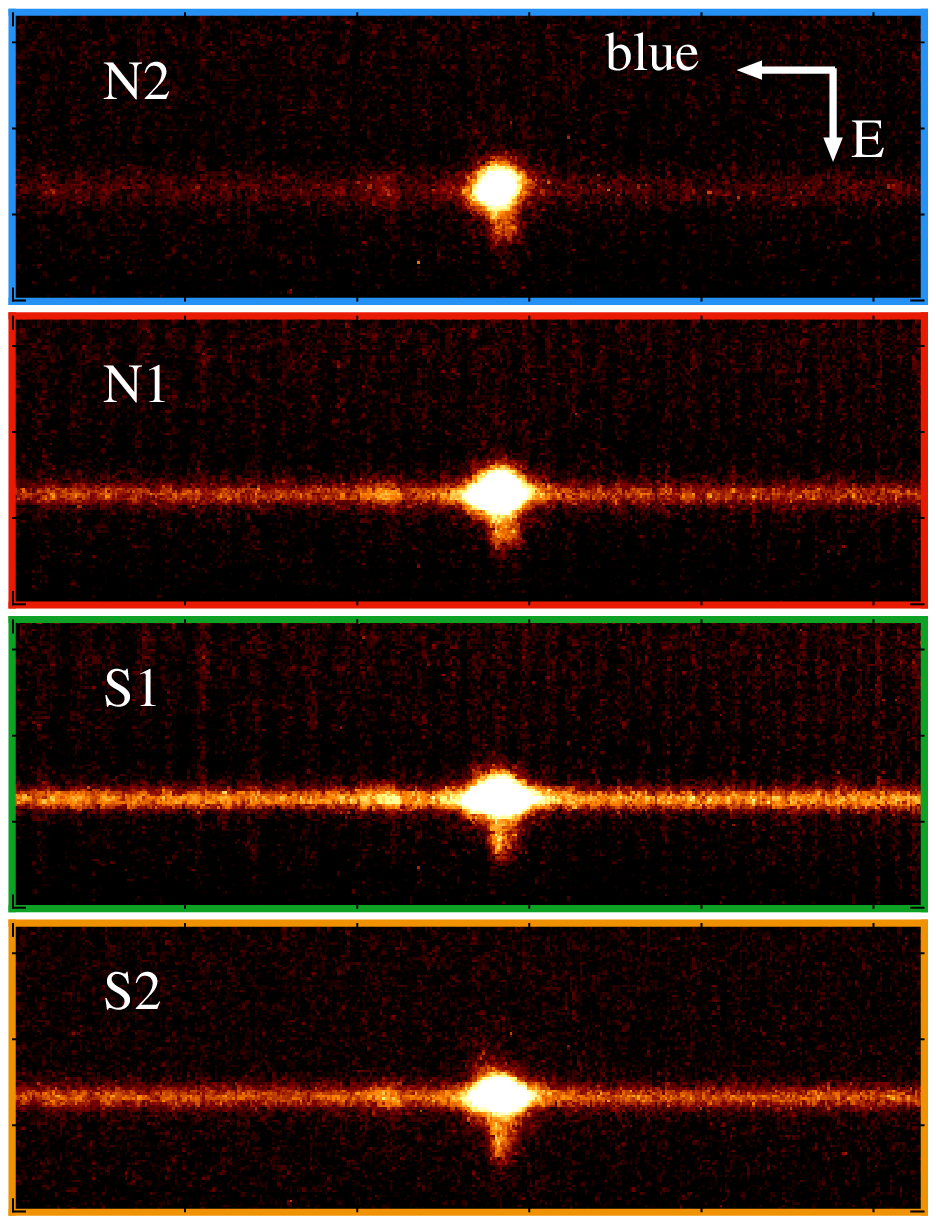}}
\caption{ Echellograms of the supernebula in \ngc . {\it Left}) K-band continuum from \scam\ showing  
slit positions.  
Contours show the 1.3 cm continuum map from \citet{turner2000}, which traces free-free radiation.  
 {\it Right})  
 In each spectrum, east runs vertically downward, and the spectral direction runs horizontally 
 with wavelength  
 increasing to the right.  
The spectra show: ({\it i}) strong \bra\ emission and continuum from the supernebula,  
({\it ii}) weaker \hei\ emission from the supernebula,  
and ({\it iii}) an extension of \bra\ emission primarily to the east of the supernebula's continuum (extending down in the figure). 0.1\arcsec\ = 1.8 pc.
\label{fig1}}
\end{figure}

\begin{figure}[!ht]
\centering
\includegraphics[width=0.6\textwidth]{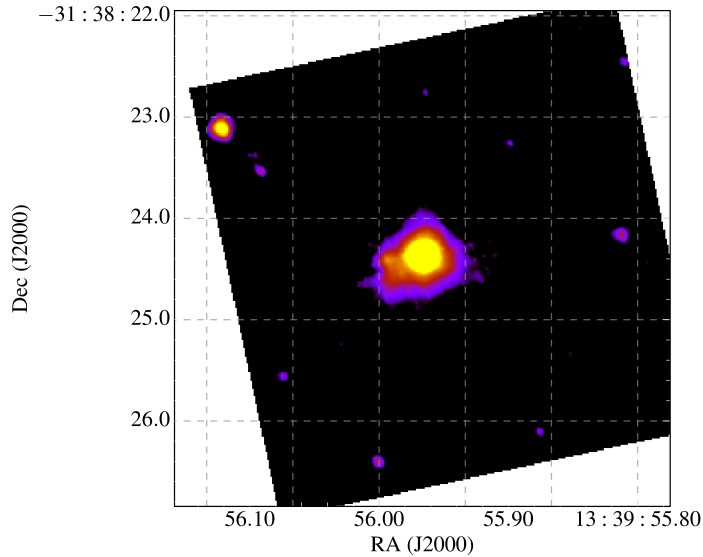}
\centering
\caption{K-band continuum of the supernebula and surrounding clusters. The slit-free, stacked \scam\ image, with resolution $\simeq$0\farcs1,
is shown here in log scale to highlight extended features. The size of the supernebula core is 0\farcs2 {\fwhm} in this image, derived 
from a Gaussian fit to the emission profile. A secondary K-band peak is detected $\simeq$0\farcs4 east of the peak emission
of the supernebula. 
\label{fig2}}
\end{figure}

The \nirspao\ echellograms
(spectrum along the $x$-axis, slit along the $y$-axis) are 
shown in Figure~\ref{fig1} along with the positions of the slits
on the stacked \scam\ image. 
The spectra
reveal strong \bra\ emission arising from the supernebula, which is spatially coincident with
the 4$\mu$m continuum.   
Blueward of \bra, there is a detection of the \hei\ 4.04899,4.04901 $\mu$m emission line doublet  \citep{hamann1986,tokunaga2000} from the supernebula.  
The \scam\ imaging, shown in Figure~\ref{fig2}, detects K-band continuum associated with
the features in the \bra\ spectra, and can be used to provide an astrometric and morphological context for the supernebula.

The {\bra} line and surrounding 4 $\mu$m continuum emission 
are strongest in slit position S1, confirming that this position is centered closest to the core of the 
supernebula. Although interpolation between slits 
to determine an exact peak is nontrivial,
the intensity variation implies a {\bra} peak that is $\lesssim$0.05{\arcsec} north of the center of the S1 slit. 
The peak of the brightest K-band source 
is likewise located between S1 and N1, 
and coincident with the {\bra} peak to within half a 0.07{\arcsec} slit.

The relative positions of the embedded radio/IR supernebula with the optical clusters in the region has been an issue of some debate,
since it is clear that there is significant extinction in the region \citep{calzetti1997,turner2003}.  
 The coordinates in HST images are not known to subarcsecond precision relative to the
 International Celestial Reference System (ICRS) coordinates, which are determined to $\lesssim$50 mas coordinates in the VLA and ALMA images of the free-free and CO(3--2) emission: $(\alpha,\delta)_{7{\rm mm}}$=(13$^{\rm h}$39$^{\rm m}$55$^{\rm s}.$9631, 31$^{\circ}$38{\arcmin}24\farcs388) \citep{turner2004}
and  $(\alpha,\delta)_{\mathrm{\co}}$=(13$^{\rm h}$39$^{\rm m}$55$^{\rm s}.$9561, 31$^{\circ}$38{\arcmin}24\farcs364) \citep{turner2017}. 
Our new AO view of the region's NIR continuum from \scam\ taken simultaneously with the Br~$\alpha$ can now tell us how the visible light and IR-radio light within the vicinity of the supernebula are related by linking the K-band   
peak, hence the bright Br~$\alpha$ source, to the optical clusters.

Registration of the \scam\ and HST images is determined by aligning the  
bright NIR clusters in the \scam\ images (some of these clusters can be seen in Fig~\ref{fig2}), with  
optical counterparts in the HST F814W image from the LEGUS survey \citep{calzetti2015}.  
Registration of the \scam\ and radio images is determined by noting that the  
bright free-free source in the radio images is the same source responsible for the brightest \bra\ emission, which we have found is coincident with the K-band peak.
 We can then align the HST images with the VLA and
ALMA images to within $\sim$50 mas. 
This registration is reinforced by visually aligning  
these registered HST images with bright clumps in the \co\ map of \citet{turner2017}; the bright \co\ peak associated with Cloud D1 is coincident with
the supernebula  and Br $\alpha$ peak. 
Aligned this way, the \co\ clouds coincide with regions of visual extinction.

\begin{figure*}[!ht]
\centering
\includegraphics[width=0.7\textwidth]{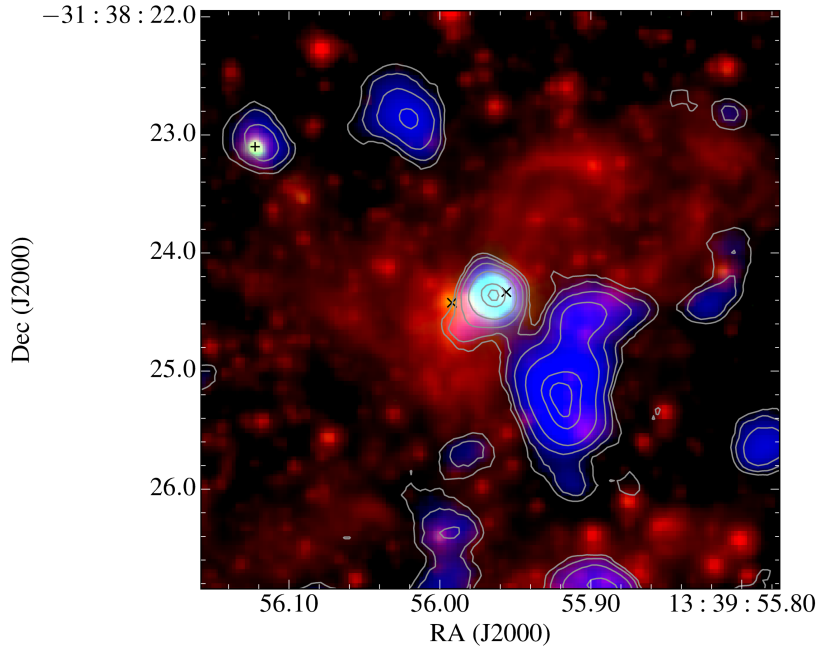}
\caption{Overlay of the K-band continuum,  
visible continuum, and CO emission near the supernebula. 
The \scam\ K-band image is in green, 
 the F814W continuum 
  from HST/LEGUS \citep{calzetti2015} is in red, and
  ALMA CO(3--2) is shown as blue and with contours \citep{turner2017}. 
  The coincidence of the NIR continuum with the \co\ cloud D1 and visible extinction 
  causes the embedded supernebula to appear light blue. 
  Positions of visible 
 clusters near the supernebula from \citet{calzetti2015} are shown as black X's: cluster \#5 and \#11 are offset 0\farcs35 east and 0\farcs14  west of the supernebula peak, respectively. 
   Cluster 105 from \citet{degrijs2013} is shown with a black cross. Positions of the optical clusters have been shifted based
  on radio registration as described in the text. 
\label{fig3}}
\end{figure*}

In Figure~\ref{fig3}, we overlay the registered
slit-free {\scam} image with the HST F814W image and the \alma\ \co(3--2) map.  
Here we identify 
visible  
clusters \#5 and \#11 from  
\citet{calzetti2015} and cluster \#105 from \citet{degrijs2013}. 
Clusters \#5 and \#11 
possibly contain stars with
ages of $\lesssim$1 Myr, based on radio fluxes and SED modeling by
\citet{calzetti2015}, who suggest 
 that these visible point-like sources could be the SSCs that power the supernebula. 
 As seen in Figure~\ref{fig3}, 
these clusters 
are not coincident with the 
K-band/\bra\ peak emission of the supernebula: 
cluster \#5 is offset by 0\farcs35 or 6.4 pc, while cluster \#11 is a closer 0\farcs14 or 2.6 pc. The compact
radio core of the supernebula \citep{turner2004} 
requires an extremely luminous 
cluster -- even
the small offset of cluster \#5 to the edge of the nebula 
would introduce a solid angle-induced luminosity augmentation that
is inconsistent with the already ``too large" cluster luminosity, given 
the dynamical mass constraints imposed by the CO linewidth \citep{turner2017}.

The perfect coincidence of the NIR continuum peak with the compact radio supernebula and  
Cloud D1, 
its extended appearance, 
and correspondence with the visible extinction, imply that 
the 2$\mu$m continuum is from dust emission. The presence of localized dust in the supernebula is supported by the detection of strong submillimeter continuum
indicating a substantial ($10^{4.2}~\rm M_\odot$) amount of dust \citep{turner2015}. If even a small fraction of
this dust is at temperatures of $T_d \sim 750~\rm K$, 
the emission is easily explained. 
Furthermore, there is excess NIR flux from the supernebula that cannot be accounted for by purely stellar+nebular models, requiring a dust emission component \citep{herrero2004,calzetti2015}. The extended nature of the K-band emission, 
with a size of 0\farcs2 {\fwhm} = 3.7 pc, suggests that the hot dust is not concentrated to the core of the supernebula cluster, especially if the eastern component (near cluster \#5) is also from dust.

Although offset east from the supernebula core, cluster \#5 is coincident with a weaker, secondary K-band peak detected with {\scam}, apparent in Fig.~\ref{fig2}. This source has also been detected in VLA radio continuum images \citep{turner2000}.
The NIR continuum emission around this component is extended out to $\sim$0\farcs3 south from its peak.
 {\bra} is also detected from the eastern component, seen in Fig.~\ref{fig1} as a tail of emission extending out to $\sim$0\farcs6 or 12 pc to the east of the supernebula peak (well past the position of cluster \#5). 
Such extended morphology calls into question whether this NIR source is indeed starlight from a cluster,
or, like the supernebula K-band source, has a significant contribution from hot dust.

In summary, our high resolution images and spectra reveal no visible clusters 
at the center of the supernebula that are likely to be the origin of the exciting UV photons. 
A strong dust continuum at 2$\mu$m 
can affect inferred properties of the SSC that rely on NIR emission lines and continuum, as noted by \citet{calzetti2015} . Specifically, measurements of age based on the equivalent widths of 
NIR recombination lines such as Br $\gamma$ become upper limits, since the contribution to the continuum from the older non-ionizing population will be overestimated. Thus, 
our observations are consistent with previous suggestions
the radiation fields  
powering the radio/IR supernebula  
are due to very young stars, 
of age $\lesssim 1$-2~Myr \citep{calzetti2015,smith2016}.

\section{BRACKETT $\alpha$ SPECTROSCOPY OF THE SUPERNEBULA} \label{sec:spectroscopy}

The \bra\ line profile and its spatial variation encode detailed gas motions, from which we can 
infer the presence of feedback and how it is affecting the SSC formation. 
Although we focus on the \bra\ line, the \hei\ 4.04899,4.04901 $\mu$m emission present in the spectra represents the first detection of this doublet in an extragalactic source, to our best knowledge. 
Other \hei\ lines have been observed in the supernebula at visible and NIR wavelengths \citep{smith2016,cresci2010}, 
indicative of massive exciting stars with high effective temperatures and  sizeable He$^+$ zones.
The echellograms suggest that the \hei\ emission is more confined to the nebular core than
 the {\bra}.  

To study the detailed line profile of \bra\ and its variation between slit positions ($\S$\ref{subsec:linewidth} and $\S$\ref{subsubsec:NSvel}), 
we first analyze integrated spectra, which are extracted by summing 12 rows$=$0\farcs215
in each echellogram centered on the peak \bra\ emission. 
The extraction box is chosen to be wide enough to detect faint features but small enough to 
probe emission from the core, matching the K-band size of the supernebula ($\S$\ref{sec:continuum}). 
The signal-to-noise at the line peak in these spectra is estimated from the rms, which includes fringing:
 ${\rm (S/N)}_{\rm peak} =$ 67, 45, 42, and 43 for slits N1, S1, N2, and S2 respectively.
We follow our analysis of integrated spectra with investigation of  
the more detailed spatial variation of \bra\ velocity as probed by 
a map of the emission, extracted by averaging spectra in bins of 6 rows$=$0\farcs108 across each echellogram ($\S$\ref{subsubsec:EWvel} and~$\S$\ref{subsec:extended}).

\begin{figure}[!ht]
\begin{center}
 \includegraphics[scale=0.48]{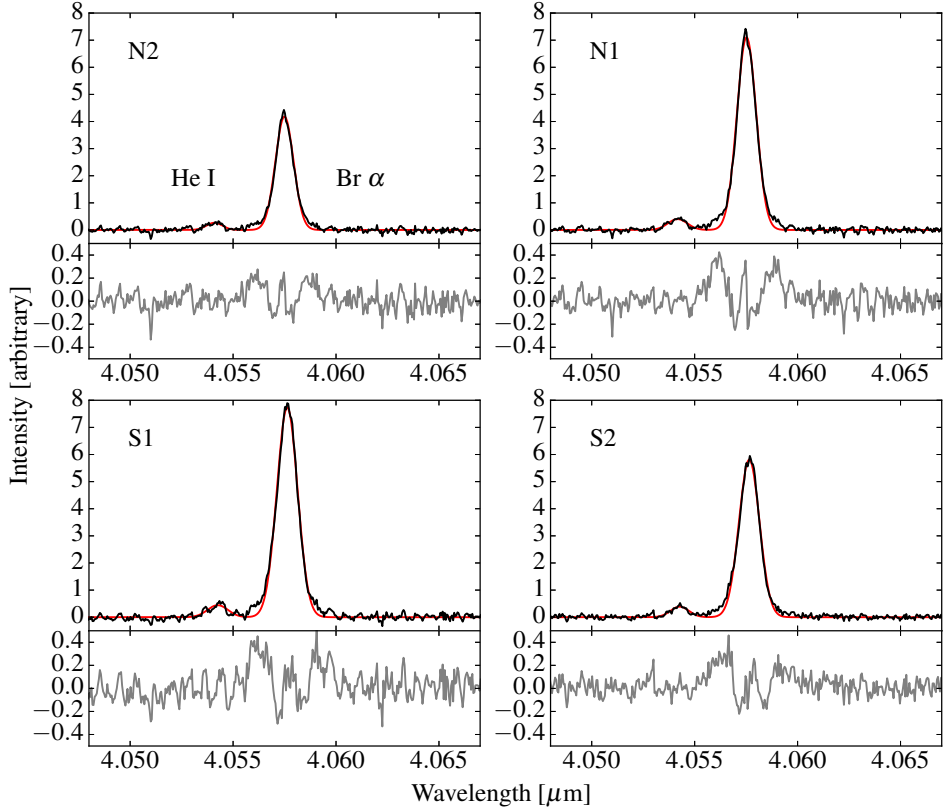}
\end{center}
\caption{ 
Continuum-subtracted, 1D spectrum extracted with a 0\farcs2 aperture from the echellograms for each of the four slit positions. For each spectrum, the top panel shows the data (black curve) and best-fit models of single Gaussian profiles for the \bra\ and \hei\ lines (red curve), and the bottom panel shows the fit residuals data$-$model (grey curve). We only fit \bra\ emission at above 10\% of the line peak to account for faint wings. 
\label{fig4}}
\end{figure}

\subsection{The Recombination Line Profile} \label{subsec:linewidth}

The \bra\ emission detected for the four slit positions across the supernebula 
 spans a range of just over 10 {\kms} in centroid velocity (about one resolution element
 of \nirspec) and exhibits 
 linewidths of {\fwhm}$=$75-87 {\kms}, based on  
best-fit models of a single Gaussian profile to emission at $>$10\% of the line peak intensity (Table~\ref{tab1} and Figure~\ref{fig4}).
The profile 
is narrower in the northern slit positions (N1 and N2), comparable with the \bra\
linewidth of $76 \pm 1$ \kms\ presented in \citet{turner2003}. In slits S1 (closest to the peak of the supernebula)
and S2, however, the line is broader by $\sim$10 {\kms}. 
 This can be compared to lower spatial resolution radio recombination lines, with 
a {\fwhm} of $58 \pm 12$ \kms\ for H53$\alpha$ \citep{rico2007}  
for a 3\arcsec\ beam, and $68\pm 3$ \kms\ for H30$\alpha$ \citep{bendo2017} for 
a 0\farcs2 beam. In the MIR, \citet{beck2012} inferred a linewidth of 65 {\kms} for the [S{\sc iv}] 10.5 $\mu$m emission within a 1\farcs2 slit. 
 
 \begin{centering}
\begin{splitdeluxetable*}{lrccccBccccc}
	\tabletypesize{\footnotesize}
	\tablecaption{\bra\ and \hei\ line properties from \nirspao\ spectra for each slit position (0\farcs2 extraction box).
	{\bra} fit parameters are quoted both for the single Gaussian model, and the two-component model consisting of
	a strong, narrow core and weak, broad wings.}%
	\tablenum{1}%
	\tablehead{
          \colhead{Slit Position}&%
          \colhead{$v_{{\rm Br}\alpha}$ [{\kms}]\tablenotemark{a}} &%
          \colhead{{\fwhm}$_{{\rm Br}\alpha}$ [{\kms}]\tablenotemark{b}}&%
          \colhead{$v_{\rm He~I}$ [{\kms}] \tablenotemark{c}}&%
          \colhead{{\fwhm}$_{\rm He~I}$ [{\kms}] \tablenotemark{d}}&%
         \colhead{$I_{{\rm Br}\alpha}/I_{\rm He~I}$ \tablenotemark{e}}&
         \colhead{$v_{\rm core}$ [{\kms}]\tablenotemark{f}}&%
         \colhead{{\fwhm}$_{\rm core}$ [{\kms}]\tablenotemark{g}}&%
         \colhead{$v_{\rm wing}$ [{\kms}]\tablenotemark{h}}&
         \colhead{{\fwhm}$_{\rm wing}$ [{\kms}] \tablenotemark{i}}&
         \colhead{$I_{\rm wing}/I_{\rm core}$ \tablenotemark{j}}
         }
\startdata
avg & $391 \pm 1$ & $84 \pm 1$ & $382 \pm 3$ & $81 \pm 10$ & $16 \pm 1$ & $391 \pm 1$ & $77 \pm 1$ & $385 \pm 5$ & $189 \pm 18$ & $0.13 \pm 0.03$  \\ 
N2 & $384 \pm 1$ & $75 \pm 1$ & $374 \pm 4$ & $67 \pm 11$ & $15 \pm 2$ & $384 \pm 1$ & $65 \pm 2$ & $386 \pm 3$ & $147 \pm 13$ & $0.22 \pm 0.05$  \\ 
N1 & $386 \pm 1$ & $80 \pm 1$ & $380 \pm 3$ & $82 \pm 8$ & $17 \pm 1$ & $386 \pm 1$ & $73 \pm 1$ & $385 \pm 3$ & $177 \pm 12$ & $0.15 \pm 0.02$  \\ 
S1 & $394 \pm 1$ & $87 \pm 1$ & $386 \pm 5$ & $86 \pm 12$ & $17 \pm 2$ & $395 \pm 1$ & $76 \pm 2$ & $385 \pm 5$ & $163 \pm 15$ & $0.20 \pm 0.05$  \\ 
S2 & $396 \pm 1$ & $86 \pm 1$ & $387 \pm 4$ & $77 \pm 10$ & $14 \pm 1$ & $397 \pm 1$ & $74 \pm 2$ & $383 \pm 5$ & $154 \pm 12$ & $0.23 \pm 0.06$  \\ 
\enddata
\label{tab1}
\tablenotetext{a}{Centroid velocity (barycentric) of the Br$\alpha$ line (from single fit). A wavelength
shift was applied to the spectra so that the \bra\ emission is at a velocity of 391 {\kms} in the average spectrum across the 
four slit positions, to match the velocity determined from H30$\alpha$ in a 0\farcs2 beam.}
\tablenotetext{b}{FWHM of \bra\ emission (from single Gaussian fit).}
\tablenotetext{c}{Centroid velocity (heliocentric) of the \hei\ line. The velocity offset of about -10 {\kms} for \hei\ relative to \bra\ could reflect our use of 4.049$\mu$m as the \hei\ rest-wavelength, while in-reality the line is a blended doublet.}
\tablenotetext{d}{FWHM of the blended {\hei} doublet.}
\tablenotetext{e}{Intensity ratio of Br$\alpha$ to {\hei}.}
\tablenotetext{f}{Centroid velocity of narrow \bra\ component.} 
\tablenotetext{g}{FWHM of the narrow \bra\ component.}
\tablenotetext{h}{Centroid velocity of broad \bra\ component.} 
\tablenotetext{i}{FWHM of the broad \bra\ component.}
\tablenotetext{j}{Peak intensity ratio of broad wings to the core.}
\end{splitdeluxetable*}
\end{centering}

\begin{figure}[!ht]
\begin{center}
 \includegraphics[scale=0.48]{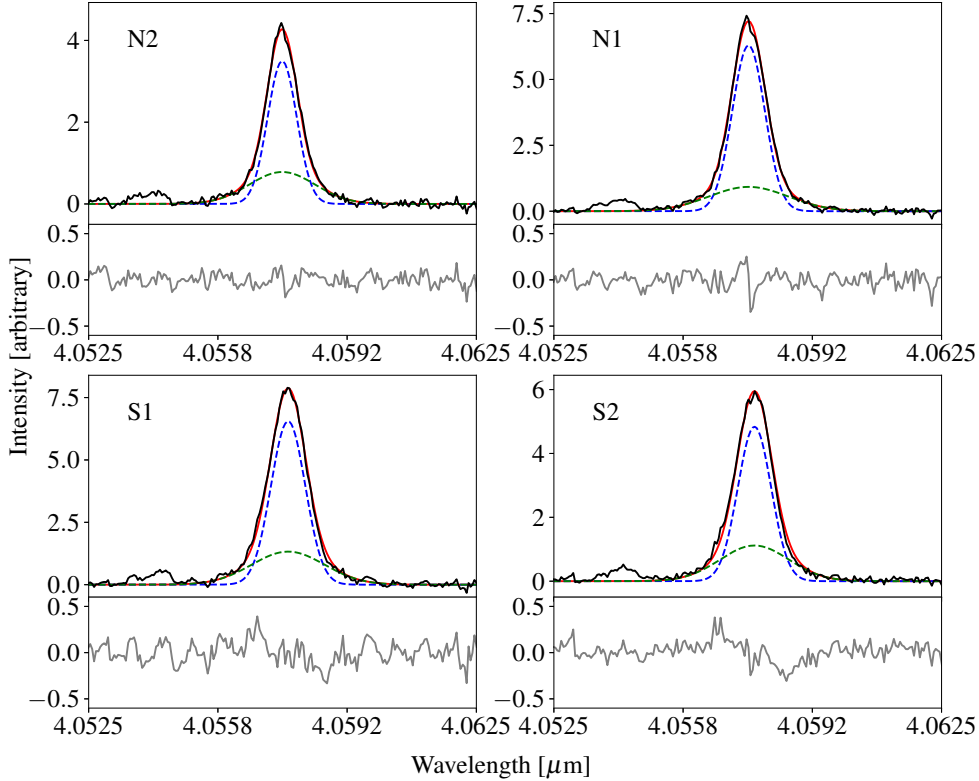}
\end{center}
\caption{ 
Same as in Fig.~\ref{fig4}, except the \bra\ is fit with a two-component Gaussian model: a weak, broad component (dashed green line) for the shoulders at the base of the line, and a stronger, narrower component corresponding to the line core emission (dashed blue). The combined fit is in red. The two-component model yields smaller residuals (compared with Fig.~\ref{fig4}) in both the wings and the very core of the line.
\label{fig5}}
\end{figure}

Figure~\ref{fig4} shows that our high-S/N spectra detect wings on the base of the \bra\ line at $\sim$10-15\% 
of the peak line intensity, which cannot be fit well by a single Gaussian model.
To account for this feature, the \bra\ profiles are additionally  
fit with a model consisting of two Gaussians: a weak, broad component and
a stronger narrow core. The resulting best-fit parameters are reported in Table~\ref{tab1}. 
The two-component model, shown in Figure~\ref{fig5}, provides a better fit in the wings and core of the observed \bra\ line, suppressing the residuals
that result from the single-Gaussian fits.   
The narrow \bra\ component is found to have {\fwhm}$ _{\rm core}=$65-76 {\kms}, smaller than
the linewidths from the single-Gaussian models, and
consistent with previous recombination line measurements which did not detect broad shoulders.  
 We discuss the implications of the \bra\ linewidths in $\S$ \ref{sec:discussion}.

In the context of cluster feedback,
the high-velocity shoulders on the base of the \bra\ emission line are very intriguing.
Our two-component models suggest broad-component widths of {\fwhm}$_{\rm wing}$$\simeq$150-175 {\kms}. Broad components with FWHM$\simeq$100-250 {\kms} have been detected on the \ha\ line in the supernebula region, but on a larger scale and for an extended gas component that is much less extincted than the supernebula gas \citep{ibero2010,westmoquette2013}. 
The \bra\ wings are roughly symmetric about the core in slits N1 and N2, 
but appear to be stronger on the blue side for slits S1 and S2, supported by a slight blueward shift in the broad component centroid in S1/S2 relative to N1/N2 (Fig.~\ref{fig5} and Tab~\ref{tab1}).
The [S{\sc iv}] emission from the supernebula exhibits excess emission extending on the blue side out to -100 {\kms} from the peak \citep{beck2012}, suggesting that the slit used for these TEXES observations could have been positioned south of the supernebula peak. However, the discrepancy in velocity is significant: a blue wing is clearly detected for the [S{\sc iv}] line and extends to high velocity, while the \bra\ wings are weak and more confined in velocity.
Nonetheless, a broad pedestal of emission appears to be a common feature for ionized gas lines in SSCs, previously observed in the Brackett lines from clusters in He 2-10 and other systems \citep[e.g.,][]{henry2007}. 
 As discussed in \S~\ref{sec:discussion}, this feature could reflect a population of individual young stellar objects (YSOs) with broad line profiles or other stars that are driving high-velocity winds \citep{beck2008}.

\subsection{Velocity Structure of the Supernebula} \label{subsec:velstruct}

The spatial variation of \bra\ emission, particularly its centroid velocity, traces the motions of gas associated with the supernebula. We can analyze the velocity structure and determine if it supports a
consistent scenario to that implied by the integrated line profile.

\subsubsection{North-South Velocity Shift} \label{subsubsec:NSvel}

\begin{figure*}[!ht]
\begin{center}
 \includegraphics[width=0.5\textwidth]{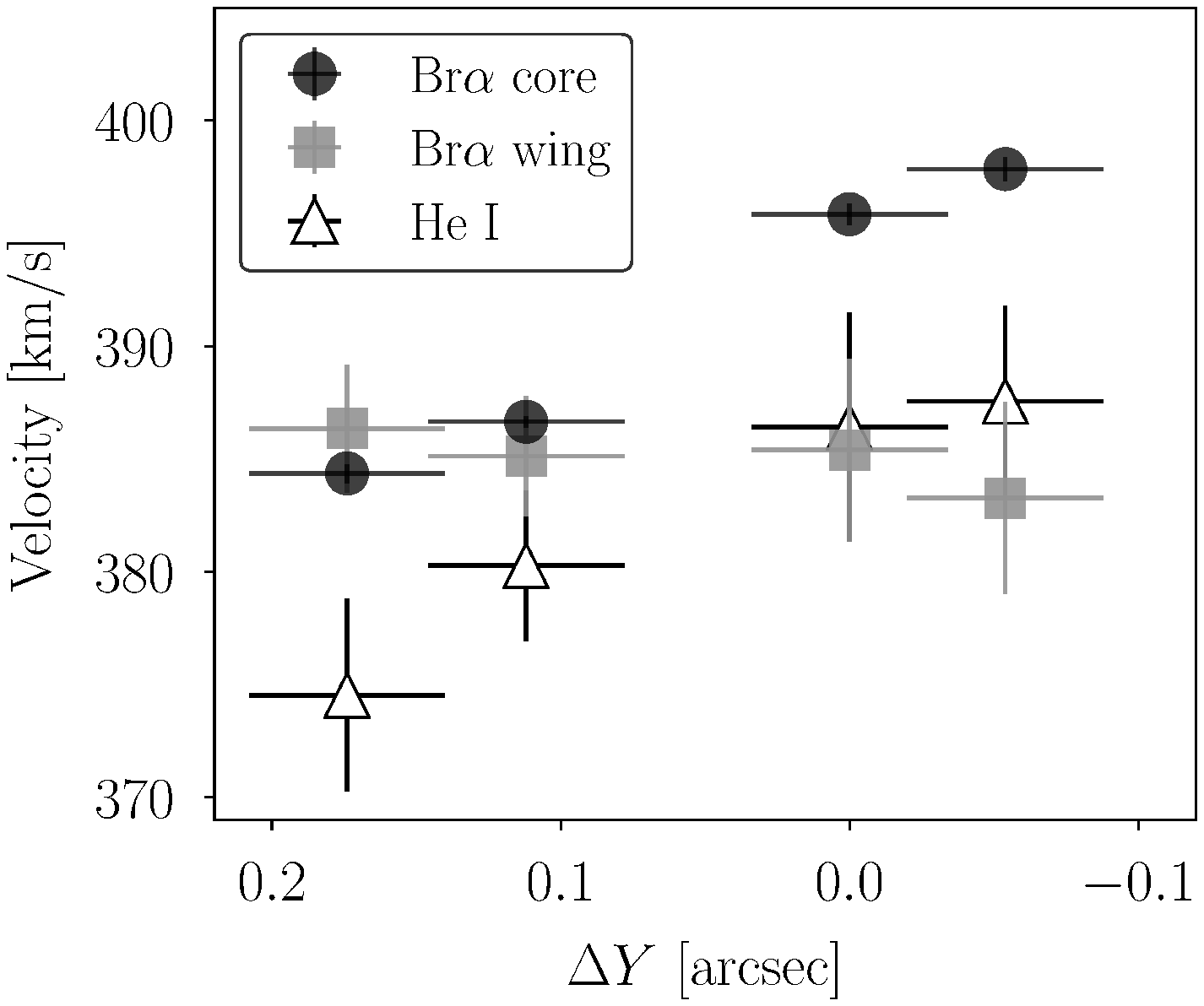}
\raisebox{9mm}{\includegraphics[width=0.48\textwidth]{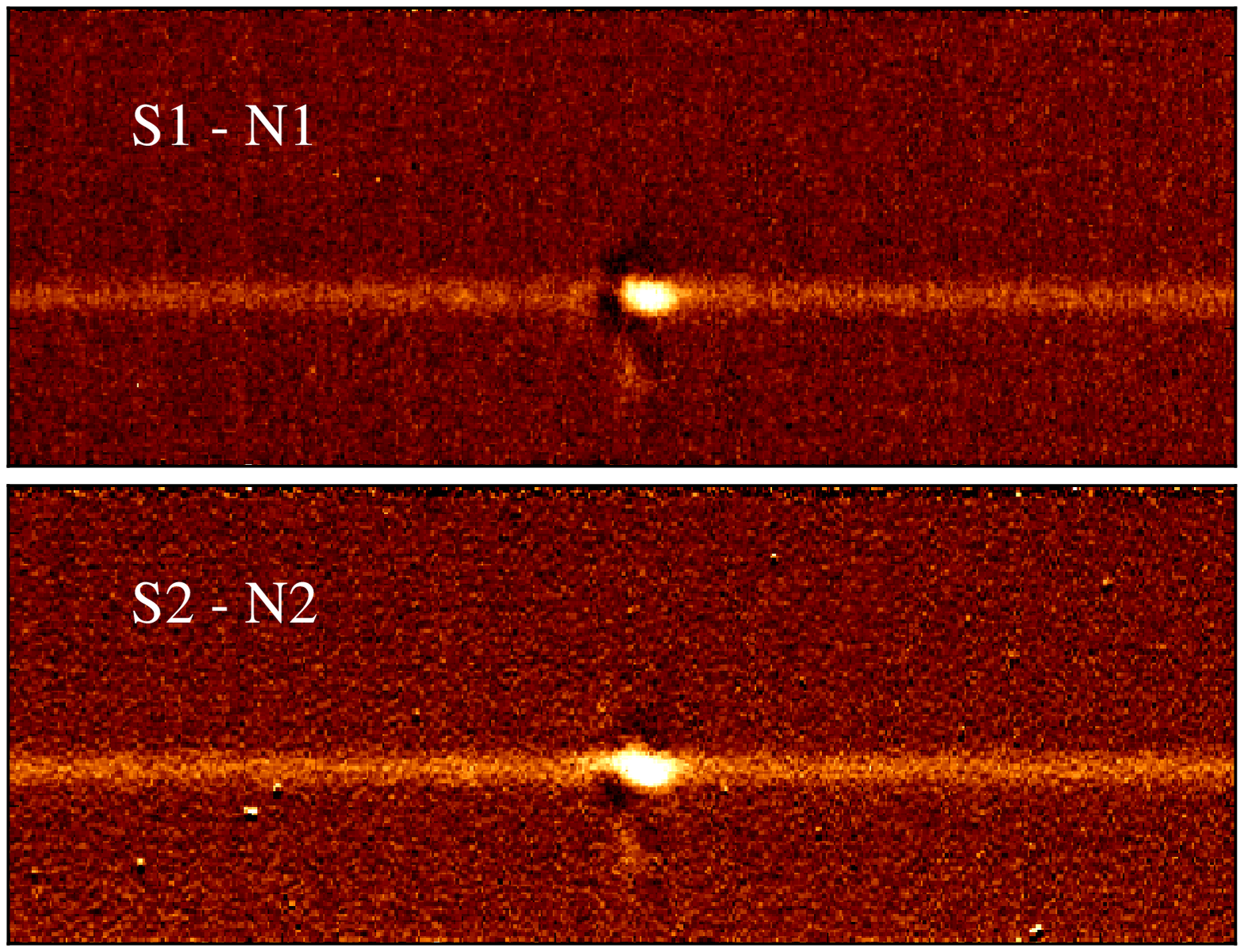}} 
\end{center}
\caption{Velocity shift across the supernebula. ~{\it Left)}: Centroid velocity (heliocentric) of the \bra\ line (narrow and broad components) and \hei\ doublet as a function of slit position in the slit-offset direction; north is left and south is right, and the zero-point is set to slit position S1. There is a clear blueshift of the \bra\ core and \hei\ doublet in the northern slit positions relative to southern by $\simeq$13 {\kms}. 
We note that \hei\ centroids are offset blueward of \bra; this likely arises from use of a single reference wavelength to derive the velocity of the \hei\ doublet. 
{\it Right)}: Difference between southern (S1/S2) and northern (N1/N2) echellograms. 
The negative arc curling around the blue side of the supernebula's
 \bra\ emission is due to the N-S line core shift 
 along with the emission in the north being more extended relative to that in the south.
\label{fig6}}
\end{figure*}

The centroid velocity  
of the supernebula core 
(Tab.~\ref{tab1}) 
exhibits a clear shift from northern to southern slit positions, shown in the left panel of Figure~\ref{fig6}.
The magnitude of this shift is +13 {\kms} across the $\sim$0\farcs3$=$5.5 pc between N2 and S2, about one full resolution element of \nirspec.
 Subtraction between the southern and northern echellograms, shown in the right panel of Fig.~\ref{fig6}, supports the \bra\ core shift inferred from the 1D spectra. The most intriguing feature in this figure is the negative arc that curls around the blue side of the southern \bra\ peak.
 This profile results from the line core shift to the blue of the northern portion, and because the northern emission is more spatially extended than the southern portion.
 Indeed, the intensity profiles along each slit indicate that the \bra\ emission is more extended by $\sim$0\farcs15 in N2 relative to that in S2.

Similar velocity structure is suggested by the H53$\alpha$ line, which exhibits a NW-SE gradient of 10 {\kms} arcsec$^{-1}$ on much larger (3{\arcsec}) scales \citep{rico2007}, along with channel maps of the H30$\alpha$ line, which reveal blueshifted emission in the NE edge of the supernebula \citep{bendo2017}. To the south of the supernebula/Cloud D1, there is another CO cloud, D4, that is redshifted relative to D1 \citep{consiglio2017}. The \co(3--2) morphology suggests a potential connection between D1 and D4 (Fig.~\ref{fig3}), which could explain the \bra\ velocity shift. Alternatively the \bra\ gradient could indicate cluster rotation, or an outflow with a bipolar morphology.
We discuss these possibilities in Section~\ref{sec:discussion}.

The velocity shifts described above apply to the narrow line core, the strongest portion of the {\bra}, but not necessarily to the broad component of the line described in Section~\ref{subsec:linewidth}. As indicated in Fig.~\ref{fig5}, the \bra\ wings are more apparent on blue side of the line for slits S1 and S2.
This could result from a slight blueshift in the velocity centroid of the broad emission from north to south, although the Gaussian fits suggest that the broad component is static across the supernebula (Fig.~\ref{fig6}). Alternatively, there could be an asymmetry of the broad component in the south, possibly owing to extinction or inherent variance between the populations of broad-line sources contained in each slit. That there are still features present
in the residuals of Fig.~\ref{fig5} supports such an asymmetry.
Ultimately, more sensitive observations of the \bra\ wings are required to study their variation near the cluster. 

\subsubsection{2D Variation in Line Core Velocity} \label{subsubsec:EWvel}

\begin{figure*}[!ht]
\begin{center}
 \includegraphics[width=0.8\textwidth]{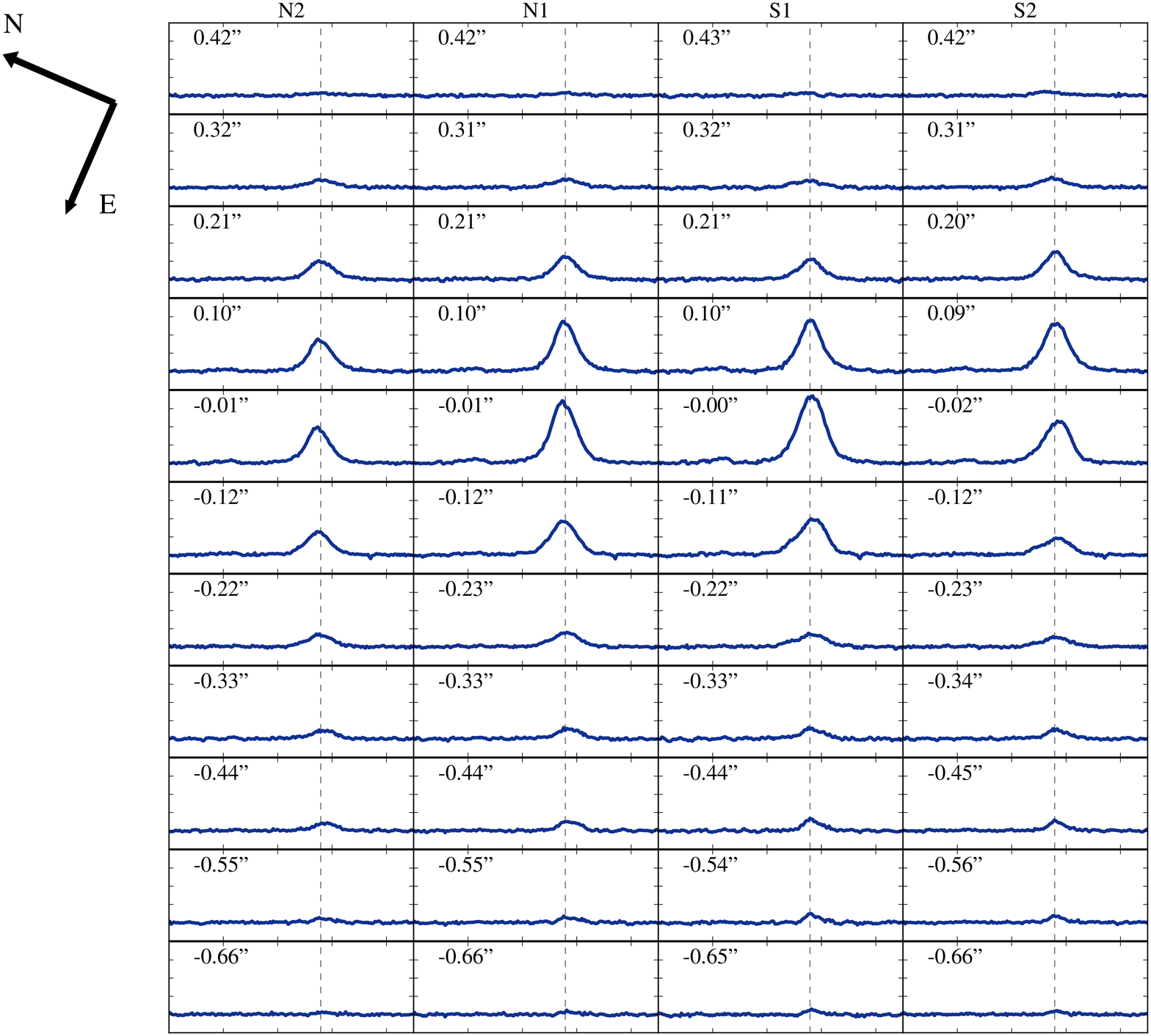}
\end{center}
\caption{ 
Map of the {\bra} line 
across the supernebula, extracted in bins of $\simeq$0.11{\arcsec} across each echellogram. Each column is a separate slit position, running north to south from left to right, and rows are such that east is down and west is up. For reference, we mark the \bra\ centroid derived for the average of all four echellograms as a dashed vertical line in each plot. Within each panel, we quote the distance from the 
peak \bra\ emission for the corresponding echellogram. 
\label{fig7}}
\end{figure*}

Figure~\ref{fig7} shows the map of the \bra\ spectrum in $\sim$0\farcs1 bins across the full $\gtrsim$1{\arcsec} extent of emission.
The map reveals the changes in line profile across the region. 
The centroid velocity of {\bra} is inferred with best-fit single-component Gaussian models to each spectrum within the map.

\begin{figure}[!h]
\begin{center}
 \includegraphics[width=0.7\textwidth]{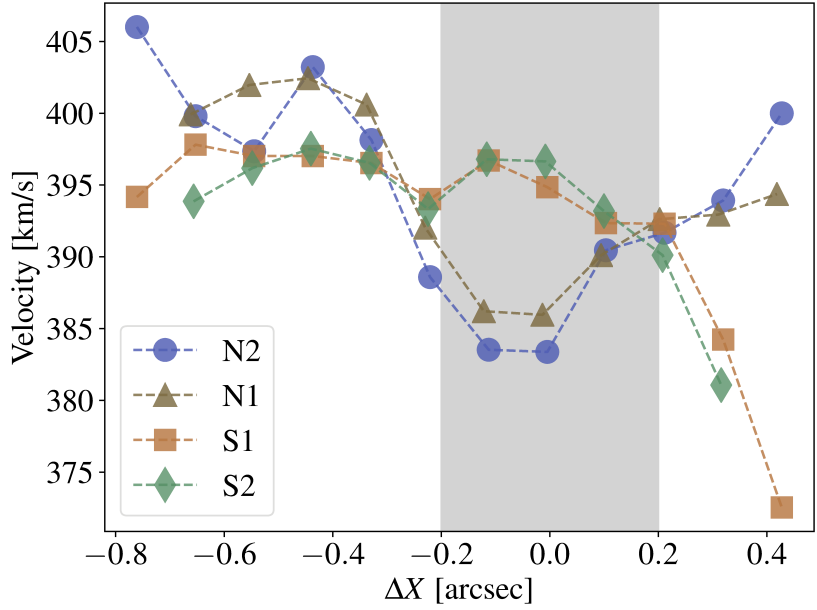}
\end{center}
\caption{ 
Centroid \bra\ velocity (heliocentric) versus position along the slit, showing the E-W velocity variation; east is left and west is right. The zero-point defines the position of the \bra\ peak along each slit. The supernebula is shaded in grey. 
\label{fig8}}
\end{figure}

The 
 \bra\ map provides a more detailed picture of the velocity structure near the supernebula, including variations of the emission
 parallel to the slits. 
Fig.~\ref{fig8} shows
 the \bra\ centroid as a function of position along the slit, for each of the four slits. 
 This plots recovers the blueshift of emission in slits N1/N2 relative to that in S1/S2, within the extent of the supernebula. 
 Moreover, for a given slit, 
  there is velocity shift from the edges the towards the center of the supernebula, 
 with the northern emission being blueshifted in the center relative to the edges while the southern emission is redshifted. This structure disfavors spherically symmetric expansion of the ionized gas.

\begin{figure*}[!h]
\begin{center}
 \includegraphics[scale=0.65]{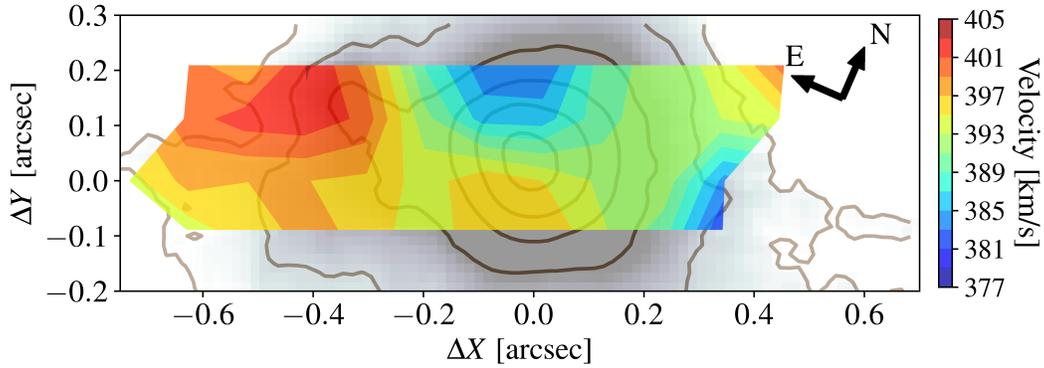}
\end{center}
\caption{ 
Map of the \bra\ velocity centroid (heliocentric) near the supernebula. The velocity map is visualized as colored, filled contours, which are overlaid on the \scam\ K-band image shown in greyscale and as brown solid contours. This figure shows the correspondence between the gas kinematics and the morphology of the supernebula. $\Delta X$ is the slit-parallel direction, with zero-point defined by the peak emission in the echellograms, and $\Delta Y$ is the slit-perpendicular position, with zero-point defined by the center of slit S1. To overlay the SCAM image, we match the K-band peak to the origin of these axes.
\label{fig9}}
\end{figure*}

The full 2D velocity structure is visualized in the context of the K-band continuum (\S~\ref{sec:continuum}) in Figure~\ref{fig9}, which overlays filled contours of centroid velocity on the \scam\ image and its intensity contours   
(akin to a moment-one map). 
The NE-SW velocity gradient across the supernebula represents the only potential detected signature of an outflow 
from the cluster, aside from the broad wings on the \bra\ line (see \S~\ref{sec:discussion}).

\subsection{Extended Emission} \label{subsec:extended}

The \nirspao\ spectra (Fig.~\ref{fig1}) show that the \bra\ emission extends $\gtrsim$10 pc (projected) to the SE 
of the supernebula's peak, well past the core radius. 
 As shown in \S~\ref{sec:continuum}, the eastern \bra\ emission is detected in the K-band imaging as a secondary peak that is coincident with the position of visible cluster \#5, surrounded by more diffuse continuum.

Gas from the eastern component is redshifted by $\sim$5-15 {\kms} relative to the supernebula, 
easily distinguished from the supernebula core (Fig.~\ref{fig9}).  
The maximal redshift of this eastern gas is concentrated to the position of the eastern K-band peak and cluster \#5. The velocity exhibits a smooth decrease with distance from away this emission peak.
That the eastern component is redshifted and has much lower extinction relative to the supernebula suggests that it is on the near side of the supernebula and infalling, with the more compact emission region (cluster \#5) leaving behind a wake of gas. Whether or not the gas in the east is related to gas to the south, which is also redshifted (Sec~\ref{subsubsec:NSvel}), is unclear.

Previous observations have found evidence for NW-SE velocity variations near the supernebula. Channel maps of H53$\alpha$ support the presence of redshifted gas extending east from the supernebula, although the lower resolution cannot identify the eastern gas as a separate component  \citep{rico2007}. Furthermore, \alma\ observations of \co(3--2) and \co(2--1) reveal a tail of molecular gas extending east from Cloud D1 on the red side of the line \citep{turner2017}. 
In the optical, a NW-SE velocity gradient is measured for a broad component and weaker narrow component of \ha, but with opposite orientation to that exhibited by \bra, such that gas is blueshifted to the east \citep{ibero2010,westmoquette2013}. However, the \ha\ is from gas associated with the optical lobes running SE-NW (Fig.~\ref{fig3}), on scales much larger than the supernebula, while the bright \bra\ is from heavily reddened gas associated directly with the supernebula.  Moreover, the \ha\ velocity gradient is most apparent for the {\it broad} component of \ha, while we only detect velocity structure for the {\it narrow} component of {\bra}. The \ha\ structure is potentially a signature of an accelerating outflow \citep{westmoquette2013}; a comparison with velocity structure for the \bra\ wings would be informative but requires more sensitive observations.

\section{The Fine Structure of Feedback in NGC 5253} \label{sec:discussion}

With recent subarcsecond observations, including those we
present here  for the
ionized gas, and for the CO and ionized gas from from ALMA 
\citep{turner2017,consiglio2017,bendo2017}, we can begin to understand the physical mechanisms occurring at the sub-cluster
scale in the forming SSC in {\ngc}.

 Our \nirspao\ observations reveal a  
 recombination linewidth of 65-76 {\kms} ({\fwhm}) in the line core, along with
 a broad wings with linewidth $\sim$150-175 {\fwhm}. The broad component is weaker, 
 constituting $\sim$25-30\% of the total line flux. The core emission is relatively narrow, considering that it is convolved
 both with the H{\sc i} thermal linewidth of 22-23 {\kms}, corresponding to an \hii\ region temperature of 11,000-12,000 K \citep{kobulnicky1997}, and the dynamical linewidth due to the cluster mass, also 21.7 {\kms}. 
 The deconvolved linewidth of $\sim$58-70 {\kms} is comparable to the linewidths of individual recombination lines
 in individual Galactic \hii\ regions \citep[e.g.,][]{jaffe1999,depree2004,sewilo2004,depree2011}.

\subsection{The Supernebula as a Compact \hii\ Region} \label{subsec:hiireg}

Feedback from \hii\ regions can be effective at disrupting GMCs and 
regulating star formation, depending on the details of their dynamical evolution and interaction with the winds from their
exciting massive stars \citep[e.g.,][]{matzner2002}. The simplest model of \hii\ region evolution, spherically symmetric
pressure-driven expansion, cannot account for the large number of ultra-compact \hii\ regions ({\uchii}) found embedded 
within massive-cluster forming regions in the Milky Way, such as in W49 \citep[e.g.,][]{dreher1984,depree2000}.

Realistic models of \hii\ region 
dynamics solve this well-known ``lifetime problem" \citep[e.g.,][]{wood1989} 
 by invoking dense, molecular gas that can prolong the expansion of the ionized region out of its compact phase. 
 For example, ``champagne flows" from blister \hii\ regions can account for much of the observed population of \uchii\ 
 and their morphology \citep{tenorio1979}. Models that can additionally explain the broad recombination lines 
 observed for a large fraction of the \uchii\ population
  invoke photoevaporation of circumstellar disks \citep[e.g.,][]{hollenbach1994,lizano1996} or mass-loading 
  of O star winds by photoevaporation of circumstellar clumps \citep{dyson1995}.
  
The \bra\ linewidth may seem small considering
the extreme cluster luminosity and youth, as noted previously 
\citep{turner2003,rico2007,beck2012,bendo2017}. 
However, if the supernebula comprises 
thousands of compact \hii\ regions within the cluster, 
the line is perfectly consistent with the line profile of an individual compact \hii\ region
 convolved with the thermal linewidth of HI and the gravitational linewidth of the 
 cluster \citep{turner2017}, as first pointed out by \citet{beck2008}.

Molecular gas, traced by CO, is
 present in the {\ngc} SSC \citep{turner2015,turner2017,consiglio2017}.
The small CO linewidth, only tracing the stellar motions due to gravity, suggests that the molecular gas is bound to individual stars. Thus, we suggest that the embedded SSC contains thousands of massive stars which are currently accreting in the form of large molecular disks, or heating molecular clumps. These O stars ionize surrounding gas and form compact \hii\ regions, which much like Galactic \hii\ regions, can be sustained at their sizes of $\sim$0.1 pc for the cluster age ($\sim$1 Myr or less) through replenishment of the expanding plasma via photoevaporation and ablation of the bound molecular gas. As discussed further in Sec.~\ref{subsec:clwinds}, radiative cooling
is likely to play a fundamental role in preventing the ionized wind regions from merging, and can
result in the formation of enriched molecular clumps.

The narrow \bra\ component suggests that most of the \hii\ regions in the SSC are expanding slowly, similar to Galactic \hii\ regions. However, the broad pedestal on the base of the \bra\ line provides a potential channel for gas escape, with linewidths that are similar to those observed around individual YSOs, which exhibit {\fwhm}$_{{\rm Br}\alpha}$=50-250 {\kms} \citep{persson1984}.
 Despite potential high-velocity winds that could be breaking out of the cluster, 
these broad-line sources contain a small fraction of the cluster's gas, and are not likely
 to drive a mass-loss rate that is presently large enough to disrupt the SSC. Nonetheless, the cluster might still 
 influence its environment through these sources by losing mass and thereby polluting the surroundings with enriched
 material.

\subsection{The Suppression of Winds \label{subsec:clwinds}}

The details of how a forming cluster disperses its gas shape its evolution and survival. Gas is expelled through winds and supernova explosions (SNe) from massive stars, and the structure and evolution of these outflows are determined by the competing effects of the outward overpressure of thermalized gas, the inwardly-directed collective gravity, and energy losses due to radiative cooling. Simulations of feedback-driven winds typically assume the winds are adiabatic, where cooling is negligible. In this case, stellar winds and SNe can merge to form a coherent cluster wind, clearing all gas from a SSC soon after formation of massive stars \citep[e.g.,][]{chevalier1985,canto2000}. In reality, radiative cooling will significantly affect the operation of mechanical feedback and substantially alter this scenario.

Cooling losses are enhanced in chemically enriched gas, and there is strong evidence that the SSC in {\ngc} cluster has enriched its environment with products of stellar evolution \citep{walsh1989,kobulnicky1997,schaerer1997,ibero2010,westmoquette2013,turner2015}. Our K-band imaging (\S~\ref{sec:continuum}) likely traces hot dust emission, localized to the supernebula and extended up to $\lesssim$10 pc to its east. Based on numerical simulations, \citet{silich2004} conclude that if the gas metallicity in the supernebula is enhanced to 1.5$Z_\odot$, winds can be entirely inhibited by cooling. \citet{turner2015} estimate that the metallicity of cloud D1 is
the $\simeq$2-3$Z_\odot$, indicating that the SSC is near or above the critical cooling regime.

Recent modeling of cluster formation find that in the most massive and dense clusters, stellar winds and even SNe around individual stars can potentially stall due to radiative cooling, and fail to merge with their neighbors \citep{tenorio2013,tenorio2015,silich2017,wunsch2017,silich2018}. 
Our \bra\ observations, suggesting that the supernebula comprises thousands of individual \hii\ regions rather than a single, merged giant \hii\ region, is consistent with critically radiative winds/SNe around the SSC's massive stars.  Indeed, {\ngc} is in the critical density regime defined in \citet{silich2017} in which it will retain its enriched stellar products due to wind stalling. The stalling mechanism can additionally explain the apparent lack of SNe signatures in the SSC without requiring an extremely young cluster age.
Another massive embedded SSC, Mrk 71-A (in the dwarf starburst galaxy NGC 2366), is similarly found to exhibit no signs of a cluster-scale outflow or superwind \citep{oey2017}.  This cluster is analogous to {\ngc}, with a mass of M$\sim$$10^5$ {\msun} and an age of $\lesssim$3 Myr. Whether the suppression of winds and resulting gas retention is standard for the most massive clusters, or whether there are other key factors such as host galaxy environment, remains a vital question in our understanding of cluster formation.

The ability of the embedded SSC in \ngc\ to hold onto its enriched gas has significant implications for the cluster's survival and potential evolution into GC. For one, continued formation of stars out of enriched material will lead to a spread in age and abundance in the stellar population, analogous to the multiple stellar populations inferred for many GCs \citep[e.g.,][and references therein]{renzini2013,piotto2015}. Moreover, the suppression of winds forces a slow expulsion of gas, preventing explosive dispersal. Removal of gas from the SSC could occur on a timescale much longer than the crossing time, allowing the cluster to respond adiabatically to changes in the gravitational potential. Along with a high star formation efficiency SFE$\gtrsim 50$\% \citep{turner2015}, the long gas removal timescale should allow the SSC to maintain most of its stars and remain bound, potentially surviving for Gyrs \citep[e.g.,][]{lada1984,kroupa2002,goodwin2006,bastian2006,baumgardt2008,allison2010,smith2013,krause2012,kim2016}.

\subsection{Origin for the Velocity Gradient Across the SSC\label{subsec:velgrad}}

In \S~\ref{subsec:velstruct}, we report a velocity gradient in the \bra\ core of $\simeq$2.4 {\kms} pc$^{-1}$ from the NE to SW edges of the supernebula, oriented perpendicular to the \nirspao\ slits. 
We suggest the following possible explanations for this feature: bulk rotation of the cluster, an outflow from an embedded source in the cluster, or a foreground infalling gas filament.

\subsubsection{Rotation of the SSC \label{subsubsec:clrot}}

Stellar dynamical studies of Galactic GCs have revealed that bulk rotation is common amongst the population \citep[e.g.,][]{kamann2018}. These investigations find evidence for a correlation between the rotation and mass of GCs, consistent with a scenario in which a forming globular cluster inherits angular momentum from its collapsing parent molecular cloud. Simulations of massive cluster formation agree with this picture, finding rotation in newborn clusters \citep[e.g.,][]{lee2016,mapelli2017}. Direct dynamical evidence for rotation exists for a handful of young massive clusters (YMCs). For example, R136 within the LMC exhibits a typical rotation velocity of $\sim$3 {\kms} based on stars within a radius of 10 pc \citep{brunet2012}. 
A velocity gradient of $\sim$1 {\kms} pc$^{-1}$ is found for the stellar surface population in the Galactic star-forming region L1688 \citep{rigliaco2016}.

The magnitude of the \bra\ shift across the supernebula is similar to those measured in YMCs for which rotation is suggested. However, the mass of the SSC in {\ngc} is orders of magnitude larger than that of such YMCs (which have M$\lesssim10^4${\msun}). The formation mechanism of SSCs is likely to differ from that of Galactic YMCs, along with the details by which angular momentum of the parent cloud is imprinted on the cluster as rotation. 
 The \co(3--2) line, associated with hot cores and individual forming stars within the cluster,  may show a slight systematic shift in its spatial centroid across the line, 
in the same sense 
as the {\bra}, with the bluer side of the line shifted about 
 0\farcs07 to the north of the red side of the line.

\subsubsection{A Bipolar Ionized Outflow \label{subsubsec:biflow}}

The \bra\ gradient could alternatively provide evidence for a bipolar outflow 
oriented along NE-SW axis. 
The implied outflow is slow, traveling at a projected speed of $\sim$10 {\kms}, likely with a true velocity of 
up to a few tens of {\kms}. Bipolar ionized outflows are commonly observed around other massive protoclusters, and can be caused by a the breaking out of winds from an embedded source, such as a massive protostellar object with an accretion disk, or a star escaping from the cluster. The outflow could even be a supernova remnant that has punched through a low density channel in the SSC. 

A similar ionized outflow is observed around the IRS2 protocluster in the Galactic star forming region W51 \citep{lacy2007,ginsburg2016}. The implied mass loss rate from this cluster is $\lesssim 10^{-3}$ {\msun} yr$^{-1}$. {\ngc} is an order of magnitude more massive than IRS2 and has an escape velocity that is correspondingly larger. At the same mass loss rate, only $\sim$2\% of the $\sim$60,000 \msun\ of gas in Cloud D1 would escape from the SSC over 1 Myr. Thus, the potential bipolar outflow 
is likely negligible in suppressing star formation, but could provide another channel through which the SSC enriches its 
environment.

\subsubsection{A Foreground Redshifted Gas Filament \label{subsubsec:infall}}

The nuclear starburst in \ngc\ is thought to be fueled by infalling filaments of cold gas, as suggested by redshifted \co\ clouds extending hundreds of parsecs along the galaxy's minor axis, associated with a prominent dust lane \citep{meier2002,turner2015}. On smaller scales there are a number of filamentary \co(3--2) clouds detected within nucleus, one or more of which could be linked to a direct flow of gas into the central SSC \citep{turner2017,consiglio2017}. Most intriguing is Cloud D4, identified in \citet{consiglio2017}. Located $\sim$10-20 pc to the south of Cloud D1 and the supernebula, Cloud D4 hosts no obvious star formation, is $\sim$3$\times$ more massive than Cloud D1, and is redshifted relative to it by $\sim$15-20 {\kms}.
As apparent in Fig.~\ref{fig3}, the \co(3--2) emission forms a bridge between clouds D1 and D4, possibly tracing gas accreting into D1 from D4.

The velocity gradient of \bra, redshifted in the SW relative to the NE, could be linked to infalling gas from Cloud D4. In this scenario the \bra\ originates in gas that is flowing from D4 to D1, ionized by the SSC. The observed gradient might then be due entirely to the gas inflow, although outflow from mechanical feedback in the NE is still possible. Unfortunately our \nirspao\ slits do not cover any positions to the south of the supernebula. Sensitive mapping of \bra\ across the region joining clouds D1 and D4, and around the other \co(3--2) clouds nearby, is necessary to provide more direct evidence of cold filament accretion.

\section{SUMMARY} \label{sec:summary}

We have obtained  0\farcs1 resolution {\nirspao} observations of the \bra\ 4.05 $\mu$m recombination line of the supernebula in {\ngc}, one of the most promising candidates for a 
young globular cluster.  Our echelle spectra (R$\sim$25,000) taken with laser-guided
AO on Keck II in four slit positions across the nebula allow for a detailed investigation of ionized gas kinematics in the region. Our findings are the following.

\begin{itemize}
\item[1.] The K-band continuum peak is found to be coincident with the {\bra} peak within $\lesssim 0\farcs035$, or 0.6 pc. 
Thus the 2$\mu$m continuum 
is coincident with radio free-free emission, the ``supernebula" along with the molecular Cloud D1 
\citep{turner2017}. The peak lies in a region of high visible extinction, and we suggest that
it is hot dust emission. 

\item[2.] The visible nuclear SSC candidates \#5 and \#11 \citep{calzetti2015}, are offset from the supernebula 
by 0\farcs35 (6 pc) and 0\farcs14 (2.6 pc), respectively. Given their separation, these sources are unlikely to power the luminous \hii\ region. Cluster \#5 coincides with a weak, secondary K-band peak.

\item[3.] The \nirspao\ spectra of the supernebula contain strong \bra\ emission, and \hei\ 4.04899,4.04901 $\mu$m emission that is $\sim$15$\times$ weaker. The
\bra\ line exhibits a small core linewidth of {\fwhm}$_{\rm core}$$=$65-76 {\kms}. 
The profile is consistent with a collection of individual (non-overlapping) compact \hii\ regions, embedded within the cluster and moving according to its gravitational potential.

\item[4.] A weak, broad pedestal is detected on the base of the \bra\ line, with a linewidth of
{\fwhm}$_{\rm wing}$$\simeq$150-175 {\kms}. This feature could trace a population of massive stars 
expelling high-velocity winds that can escape the SSC.

\item[5.] 
The \bra\ emission is extended to the east of the supernebula, near 
cluster \#5, and 
is redshifted by $\simeq$5-15 {\kms} relative to the supernebula. 
The extended gas is likely foreground to the supernebula and falling towards it.
 It remains unclear whether cluster \#5 is indeed a star cluster which is in the process
of merging with the central SSC, or a dense knot of gas/dust that reflects the visible light of the SSC and may 
be enriched with material that has been expelled from it.

\item[6.] The centroids of the narrow \bra\ component and the \hei\ doublet exhibit a velocity shift of +13 {\kms} from the northeast to southwest edge of the supernebula. A similar velocity shift of smaller magnitude is seen in \co. 
The velocity profile is inconsistent with spherical expansion/outflow, but could be due to: rotation along the axis parallel to the slits, a biconical outflow from an embedded source breaking out of the cluster with a velocity of $\sim$10-50 {\kms}, or accretion of gas from a massive, redshifted molecular cloud to the south of Cloud D1.

\end{itemize}

We suggest that we see in the supernebula/Cloud D1 region the dynamics of individual ultra-compact \hii\ regions around massive stars within the giant cluster that powers the supernebula.
 Winds and supernovae from these massive stars may be stalled due to critical radiative cooling, and 
cannot merge to generate a cluster-scale superwind.
We detect two possible sources of outflow: the broad component of \bra\ along with the velocity gradient of the narrow component across the supernebula. Neither of these appear to be presently capable of rapidly removing a 
large amount of gas from the SSC.
While {\ngc} has been thoroughly studied, its context in the general formation of massive clusters (such as GCs) remains unclear. Is the embedded SSC unique in its lack of high-velocity gas dispersal, or is it typical of SSCs of a given mass along their evolutionary paths? Further high-resolution, infrared spectroscopic studies of forming massive clusters can probe ionized gas to sub-cluster scales, and peer past the veil of dust in which the stars of a young cluster are embedded.

\acknowledgments

We thank Dr.~Sergiy Silich (Instituto Nacional de Astrof\'{i}sica, \'{O}ptica y Electr\'{o}nica) and the anonymous referee
for valuable discussion and recommendations for the manuscript. The data presented herein were obtained at the W. M. Keck Observatory, which is operated as a scientific partnership among the California Institute of Technology, the University of California and the National Aeronautics and Space Administration. The Observatory was made possible by the generous financial support of the W. M. Keck Foundation. The authors wish to recognize and acknowledge the very significant cultural role and reverence that the summit of Maunakea has always had within the indigenous Hawaiian community.  We are most fortunate to have the opportunity to conduct observations from this mountain.

\facility{Keck:II (NIRSPAO)}
\software{IRAF, SAOImage DS9, Astropy \citep{astropy} }

\bibliographystyle{aasjournal}
\tracingmacros=1
\bibliography{n5253_nirspao}

\end{document}